\documentclass[aps,prx,twocolumn,showpacs,amsmath]{revtex4-1}

\usepackage{color}
\usepackage{graphicx}
\usepackage{epsfig,psfrag}
\usepackage{amsmath}
\usepackage{amssymb}
\usepackage{amsbsy}
\usepackage{amsthm}
\usepackage{amsfonts}
\usepackage{wasysym}
\usepackage{bbm}
\usepackage{tabularx}
\usepackage{euscript}
\usepackage{color}
\usepackage{enumerate}
\usepackage{amsfonts}
\usepackage{exscale}
\usepackage{bbold}
\usepackage{float}

\usepackage[colorlinks,citecolor=blue]{hyperref}

\renewcommand{\v}[1]{{\boldsymbol{#1}}}

\newcommand{\be}{\begin{eqnarray}}
\newcommand{\ee}{\end{eqnarray}}

\newcommand{\Eq}[1]{equation~(\ref{#1})}
\newcommand{\Fig}[1]{Fig.~\ref{#1}}

\newcommand{\nn}{\nonumber\\}

\renewcommand{\>}{\rangle}

\newcommand{\s}{{\sigma}}
\def\bea{\begin{eqnarray}}
\def\eea{\end{eqnarray}}

\def\avg#1{\left\langle#1\right\rangle}

\def\Eq#1{Eq.~(\ref{#1})}
\def\Fig#1{Fig.~\ref{#1}}

\begin{document}

\title{Superconductor to metal transition in overdoped cuprates}
\author{Zi-Xiang Li$^{1,2}$
Steven A. Kivelson$^{3}$,}
\author{Dung-Hai Lee$^{1,2}$} \thanks{dunghai@berkeley.edu}
\affiliation{
$^1$ Department of Physics, University of California, Berkeley, CA 94720, USA.\\
$^2$ Materials Sciences Division, Lawrence Berkeley National Laboratory, Berkeley, CA 94720, USA.
}
\affiliation{
$^3$ Department of Physics, Stanford University, Stanford, California 94305, USA.}

\begin{abstract}
We present a theoretical framework for understanding the
behavior of the normal and superconducting states of overdoped cuprate high temperature superconductors in the vicinity of the doping-tuned quantum superconductor-to-metal transition. The key ingredients
on which we focus are $d$-wave pairing, a flat antinodal dispersion, and disorder.  Even for homogeneous disorder, these lead to effectively granular superconducting correlations and a superconducting transition temperature determined in large part by the superfluid stiffness rather than the pairing scale.
\end{abstract}

\maketitle

\section{Introduction}
For over three decades, research on the cuprate superconductivity primarily focused on the underdoped and optimally doped region of the phase diagram.
Here, it is  now  widely accepted that $T_\text{c}$ is not set by the scale of Cooper pairing (as in BCS theory), but is instead largely determined by the onset of phase coherence (i.e. by the superfluid density)\cite{Uemura-1989,Kivelson-1995}.
Phenomena such as the pseudogap, intertwined orders, and strange metal behavior remain the focus of
considerable research today. In contrast, it is commonly believed that the physics of the overdoped cuprates is more conventional. For example, angle-resolved photoemission spectroscopy (ARPES) shows a large untruncated Fermi surface in the normal state with reasonably well-defined quasiparticle peaks, and a superconducting gap that decreases with increasing doping, more or less in tandem with  $T_\text{c}$\cite{Shen-2003}.
Moreover, in at least one material\cite{Hussey-2008}, quantum oscillations, of the sort expected on the basis of band-theory, have been documented.

Thus, it
was a surprise that recent penetration depth measurements\cite{Bozovic-2016,uemura2} on  crystalline LSCO films suggest that the superconductivity in the overdoped cuprates is also limited by the onset of phase coherence.
Consistent with this result, recent ARPES measurements\cite{ZXnew} of overdoped Bi2212 found spectroscopic evidence that Cooper pairs are already formed at temperatures about 30\% higher than $T_\text{c}$. Adding to the puzzle, recent optical conductivity measurements\cite{Armitage-2019} showed that below $T_\text{c}$ a large fraction of the normal state Drude weight remains uncondensed. This is consistent with
earlier specific heat measurements
which show a
 $T$-linear term that persists to the lowest temperatures, $T\ll T_\text{c}$, with a magnitude that is a substantial fraction of its normal state value\cite{Wen-2004,TallonTl2201}.
A possibly related observation\cite{Davis-2008,gomes,Yazdani-2008}  from scanning tunneling microscopy (STM) is that, at least up to moderate levels of overdoping, a spectroscopic gap persists in  isolated patches up to temperatures well above $T_\text{c}$, so that the normal state electronic structure is suggestive of superconducting grains embedded in a normal metal matrix.
Other than the STM results (for which the relevant data do not exist at very high overdoping),  these  phenomena become increasingly dramatic as the doped hole concentration, $p$, approaches the critical value, $p_{\rm smt}$, at which the superconductor-to-metal transition occurs at the overdoped end of the superconducting dome.

The primary goals of this paper are to present a simple theoretical model that captures 
 what we believe to be the essence of the above phenomena, and to explain the cause of the superconductor-metal transition.  We are aware that our model does not capture various quantitative aspects of the actual materials. In the next two paragraphs we present the physical picture underlying this work.

\begin{figure}[h]
	\begin{center}
		\includegraphics[scale=
		0.3]{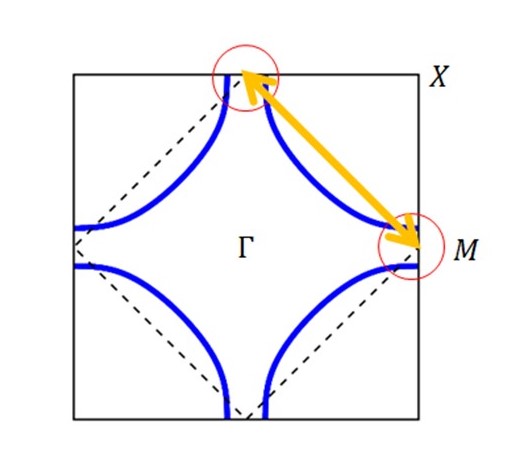}
		\caption{
		\textbf{Model Fermi surface}.  Antinode to antinode scattering induced by the disorder potential is indicated by the yellow arrow. }
		\label{saa}
	\end{center}
\end{figure}

A sketch of the Fermi surface  of 
an overdopped cuprate is shown in  \Fig{saa};  it is shown as being hole-like, although in
some  cuprates the Fermi surface passes through a Lifshitz transition at a doping concentration, $p_{\rm Lif} < p_{\rm smt}$, in which case it would be electron-like\cite{Yoshida-2001,Valla-2018}.  The neighborhood of the van-Hove points - which we will refer to as the anti-nodal regions - is also the portion of the Fermi surface  farthest from the gap nodes in the d-wave superconducting state and so is where the gap is largest. As we will discuss, the fact that the Fermi surface passes near the van-Hove point, meaning that the Fermi velocity is small in the anti-nodal regions, plays a significant role in the results we obtain; whether it is electron or hole-like is relatively less important.
In considering the effects of disorder, the scattering between anti-nodal regions (indicated by the orange line in the figure) is particularly important, as it is pair-breaking. Such anti-node to anti-node scattering
is apparent in  STM quasiparticle inteference measurements in both non-superconducting \cite{Wang-2018} and superconducting\cite{Hoffman-2014} overdoped Bi2201. In particular in Ref.\cite{Hoffman-2014} it is shown that at voltages corresponding to the anti-nodal gap energy, the Fourier transform of the local density of states exhibits a broad maximum at momentum $\v q=(\pi,\pi)$. Moreover, from the coherence factor, it is inferred that this scattering occurs between momentum regions having opposite signs of gap the  function\cite{ Hoffman-2014}.

When the Cooper pair coherence length is comparable to the correlation length of the disorder potential, the prior mentioned pair-breaking causes the pair field amplitude to be spatially heterogeneous\cite{Kivelson-2008,Kivelson-2018}. The superfluid stiffness is large in regions with high pair field amplitude, whereas the stiffness is low where the pair field amplitude is small. This is reminiscent of a granular superconductor. The small stiffness in the inter-granular regions causes the averaged zero temperature superfluid density to be low, hence
 superconducting phase fluctuations (both classical and quantum) are enhanced.
 In the metallic regions the Cooper pairing instability is inhibited since a) the repulsive interactions between electrons
 force the average of the superconducting order parameter to be zero around the Fermi surface, and b) a sign changing order parameter is suppressed by disorder scattering.
 The un-paired electrons in the inter-granular regions give rise to a substantial uncondensed Drude component in the optical conductivity and
 to a residual $T$ linear term in the specific heat. A superconductor-to-metal transition occurs when the superconducting islands grow sufficiently sparse\cite{Kivelson-2008}.

Note that the normal state transport is dominated by the nodal quasiparticles, which are known to be less affected by  impurity scattering \cite{ZXShen-2005,Hanaguri-2007}.
In the rest of the paper we present results corroborating the physical picture presented above.\\

\section{Results}
\subsection{The model}

The model we use to describe the superconducting state contains 
hopping terms, interaction terms, and disorder potential terms.
The Hamiltonian is
\be
H&&=
-\sum_{i,j,\s}t_{ij}\left(c^\dagger_{i\s}c_{j,\s}+h.c.\right)\nn
&&+\sum_{i,\s} (w_i -\mu) c^\dagger_{i\s}c_{i,\s}
+H_{\rm int}
\label{model}
\ee
where $t_{ij}$ is the hopping integral between sites $i$ and $j$ on a square lattice which we take (to produce a cuprate-like Fermi surface) to be $t_{ij}=1$ between nearest-neighbor sites, $t_{ij}=-0.35$ between second-neighbor sites, and $t_{ij}=0$ for all further neighbors. To compare different pairing symmetries, we consider two different forms of $H_{int}$: 1) As a model of a d-wave superconductor (relevant to the cuprates) we adopt a model with a nearest-neighbor antiferromagnetic Heisenberg exchange interaction,
 $H_{\rm int}=J\sum_{\<ij\>}\v S_i\cdot\v S_j$. 2) As a model of an s-wave superconductor, we consider an attractive Hubbard interaction,
 $H_{\rm int}=-U\sum_{i}c^\dagger_{i\uparrow}c_{i\uparrow}c^\dagger_{i\downarrow}c_{i\downarrow}$.
We fix the strength of the pairing interactions to $J=0.8$ and $U=1.35$, respectively, so
that the two superconducting gap scales in the absence of disorder are approximately the same.

In treating the problem with disorder, we consider a finite system of size $40\times 40$ and, unless otherwise indicated, assume periodic boundary conditions. The random potentials, $w_j$, represent the effects of disorder:  on a randomly chosen fraction $n_{\rm imp}$ of sites we set $w_j=w>0$, with $w_j=0$ on all other sites. Modelling disorder as point-like impurities is a simplification. In reality the potential produced by the dopants can extend over multiple unit cells. The point-like impurity is considered because it can cause the large momentum transfer anti-node to anti-node scattering seen in STM\cite{Wang-2018, Hoffman-2014}, which is a key ingredient in our theoretical framework. In the main text we report results for disorder strength, $w=1$, but in the Supplementary Note 2 we include results for other values - the main qualitative results do not depend sensitively on the value of $w$.
We repeat this with multiple different impurity configurations in order to compute the configuration averages of physical observables;
typically, we  average over $64$ distinct impurity configurations
 but we average over $128$ configurations when the impurity concentration is large and the superconducting pairing is highly inhomogeneous.

\begin{figure}[t]
	\begin{center}
		\includegraphics[scale=0.35]{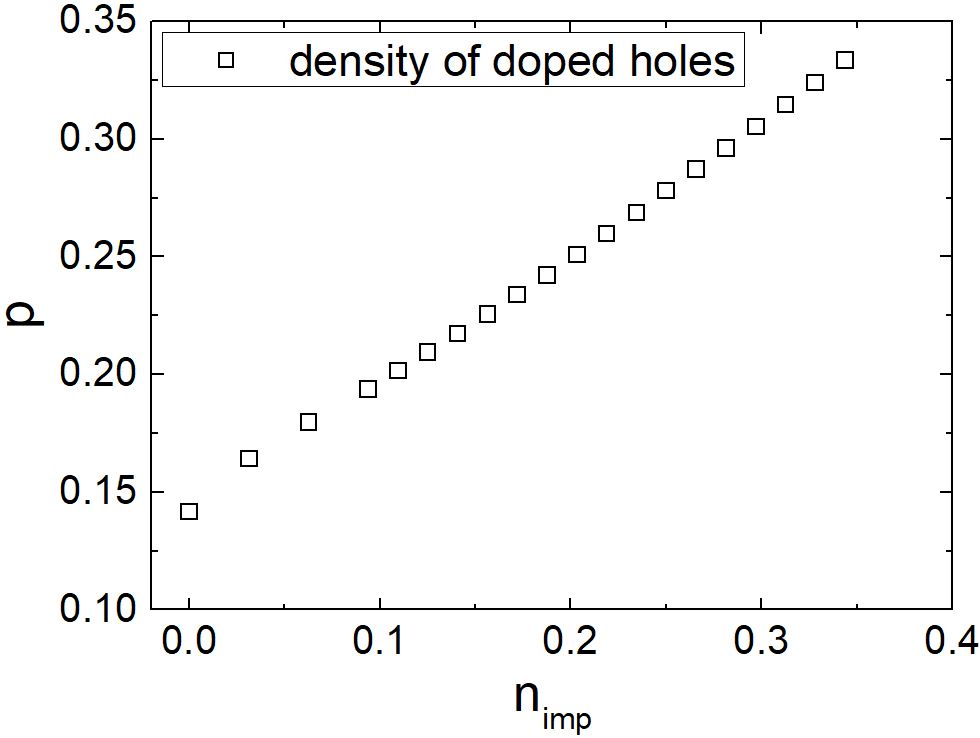}
		\caption{\textbf{Plot of the density of doped holes $p$ as a function of impurity concentration $n_{\rm imp}$ with $w=1$ }. }
		\label{dopinglevel}
	\end{center}
	\end{figure}

We solve the model by self-consistent BCS mean-field approximation. However, the disorder scattering is treated exactly. Since
$H$ lacks translation invariance,  the self-consistency equations need to be solved numerically. Since our calculation does not capture the thermal and quantum fluctuations, we focus
 primarily on zero temperature and on doping sufficiently away from the quantum critical doping of the superconductor-metal transition.

The carrier concentration is controlled by the chemical potential $\mu$ and the impurity concentration $n_{\rm imp}$. Sometimes we fix $\mu$ while changing $n_{\rm imp}$ to reach the desired carrier (hole) concentration. In the case where we want to study the effect of disorder at a fixed carrier density we tune $\mu$ while varying $n_{\rm imp}$ to achieve the desired carrier density. To avoid the complex issues of the pseudogap, intertwined orders and strange metals, we take our lowest carrier concentration to be slightly larger than optimal doping. Thus except in \Fig{flatdisp}, the impurity concentration is measured relative to that present at optimal doping. For example, in \Fig{dopinglevel} we show the hole concentration $p$ as a function of $n_{\rm imp}$ at fixed $\mu$ with $w=1$.

\subsection{The mean-field solution}

\begin{figure}[t]
	\begin{center}
		\includegraphics[scale=0.6]{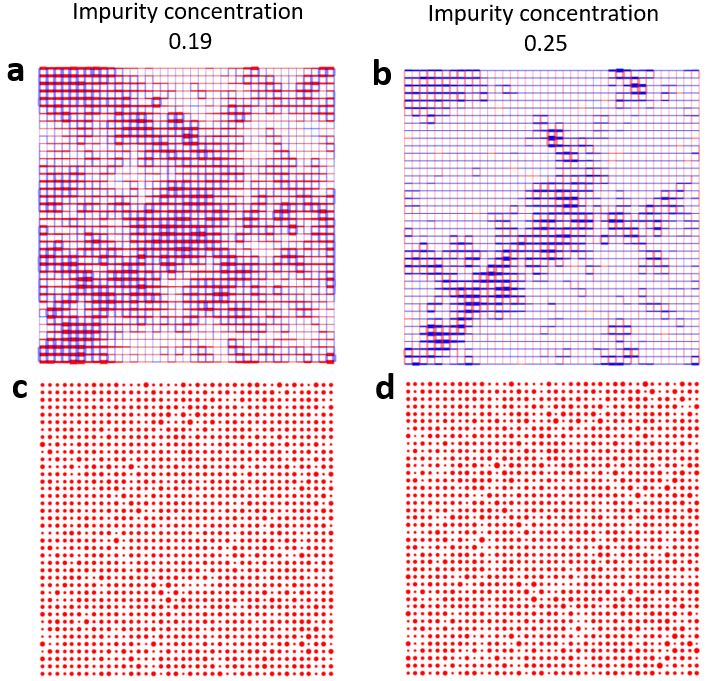}
		\caption{\textbf{The real space distribution of the pair field}. The upper two panels are for $d$-wave pairing (where the pair fields lie on nearest-neighbor bonds);
		the lower two panels are for $s$-wave pairing (where the pair fields are on-site).  The size of the symbols, namely the thickness of the bonds in panel \textbf{a} and \textbf{b}, and the size of the dots in panel \textbf{c} and \textbf{d}, represents the magnitude of the pair field whereas the color (red positive, blue negative) the sign. The left and right columns correspond to two different impurity concentrations.
		The magnitude of pair field
		in panel a ranges from $0.0003$ to $0.103$ while that in panel b ranges from $0.000005$ to $0.1008$. In panel \textbf{c} and \textbf{d}, the magnitude of pair field on each site
		ranges from $0.076$ to $0.18$ and $0.073$ to $0.20$, respectively.
		}
		\label{realspace}
	\end{center}
\end{figure}

\begin{figure}[t]
	\begin{center}
		\includegraphics[scale=0.4]{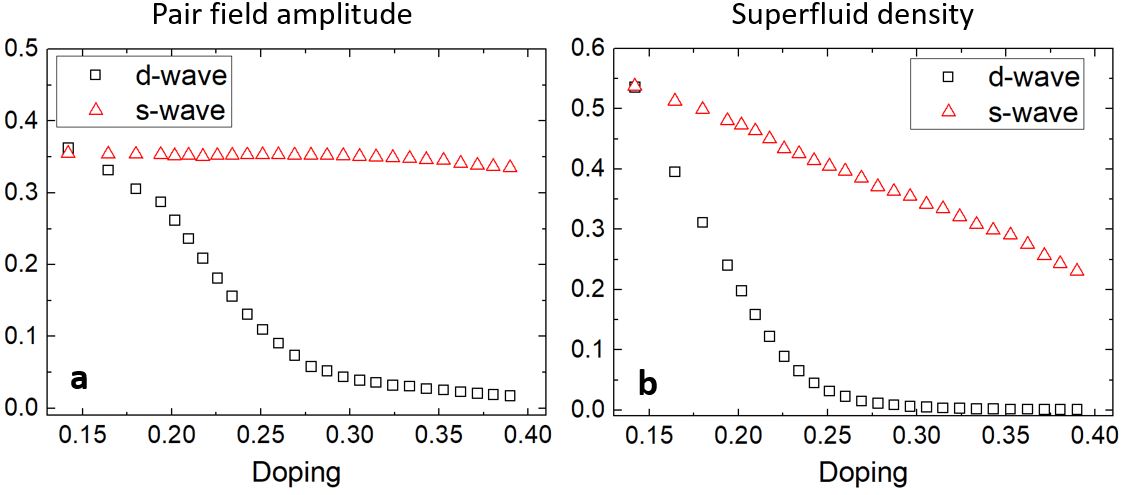}
		\caption{\textbf{The doping dependence of the spatial averaged zero-temperature
		magnitude of the pair field and the mean-field value of the superfluid density}. The band-structures used in these plots are the same, the only difference is the pairing interaction. The red symbols represent $s$-wave pairing while the black symbols represent the $d$-wave pairing. In the band-structures used in constructing these plots, the Fermi surface crosses van-Hove point, namely Lifshitz transition occurs, at the doping level $p_{\rm Lif} \approx 0.33$.}
		\label{pairingamp}
	\end{center}
\end{figure}
The local value of the gap parameter, $\Delta_{ij}$, that enters the mean-field equations, which we will refer to as the pair field,  is given by the product of the pairing interaction times the expectation value of the pair  annihilation operator.  In the  s-wave case, $\Delta$ is site diagonal,  $\Delta_j\equiv \Delta_{jj}= U \langle c_{j\uparrow}c_{j\downarrow}\rangle$ while for the d-wave case, $\Delta_{ij} = J \langle c_{i\uparrow}c_{j\downarrow} + c_{j\uparrow}c_{i\downarrow} \rangle$, where $i,j$ are any pair of nearest-neighbor sites.  The self-consistently computed values of the pair field for two impurity configurations with different doped hole concentrations are shown in Fig. \ref{realspace}.

In the s-wave case, we find that $\Delta_j$ has a uniform sign and  a magnitude that is weakly dependent on position. Moreover, it does not depend on the doped hole concentration strongly. The red symbols in Fig. \ref{pairingamp}a show the configuration averaged value of $|\Delta_i|$,  as a function of $p$.

In the d-wave case, $\Delta_{ij}$ has a magnitude that varies significantly as a function of position and which is strongly doping dependent.  It also reflects the d-wave symmetry of the uniform state from which it descends in that, with minor exceptions, $\Delta_{ij}$ is positive on  bonds oriented in the $\hat x$ direction, and negative on  bonds in the $\hat y$ direction.   The black symbols in  Fig. \ref{pairingamp}b show the configuration average of $|\Delta_{ij}|$ as a function of $p$.  Notice that it drops dramatically with increasing $p$, but then has a long tail with small magnitude that extends to high values of $p$.

Note that the band structures used in the two cases are the same; only the pairing interaction is different. The
dramatic contrast between the two cases manifests Anderson's theorem for the $s$-wave, and pair breaking by the scalar
disorder  for the $d$-wave.

Specifically, in the d-wave case, pair-breaking induced by the anti-node to anti-node scattering tends to strongly suppress
 superconducting pairing. A consequence of this is that when disorder is strong, the pair field amplitude becomes granular (heterogeneous) with significant pairing occurring only in rare regions where disorder is weak\cite{Kivelson-2018}. This can be seen clearly in Fig. \ref{realspace}b. The superconducting order parameter on different grains are connected by effective SNS (superconductor - metal - superconductor) Josephson junctions
  which, as suggested in \cite{Kivelson-2015}, can vary randomly in sign due to the sign-changing superconducting order parameter. This can cause frustration in the superconducting phase coherence, and ultimately, as we will see shortly, to spontaneous time-reversal-symmetry breaking and the existence of local super-current loops.
     \\

The net superfluid density is computed from the standard Kubo formula,
Supplementary Eqs. 2 and 3.  In \Fig{pairingamp}b we compare the $p$ dependence  of the  $T=0$ mean-field  superfluid density  for the $s$-wave and $d$-wave pairing cases.  Notice that for the d-wave case, the superfluid density drops considerably more rapidly than does the pair field amplitude.  (This can be seen more quantitatively in  Supplementary Figure 1, where the d-wave case is shown on a log-linear scale.) This implies that phase fluctuation effects, beyond the mean-field treatment, must inevitably become large in this range of doping. \\

We have also computed the $T$ dependence of the mean-field superfluid density.  For the d-wave case, the results are shown in Supplemental Figure 4.  At low $T$ (where we see a $T$-linear decrease), the results may be physically meaningful, but at higher temperatures thermal phase fluctuations, which
are ignored in our mean-field treatment,  must certainly play a role in the vanishing of the superfluid density as $T\to T_\text{c}$.
\\

\begin{figure}[t]
	\begin{center}
		\includegraphics[scale=0.38]{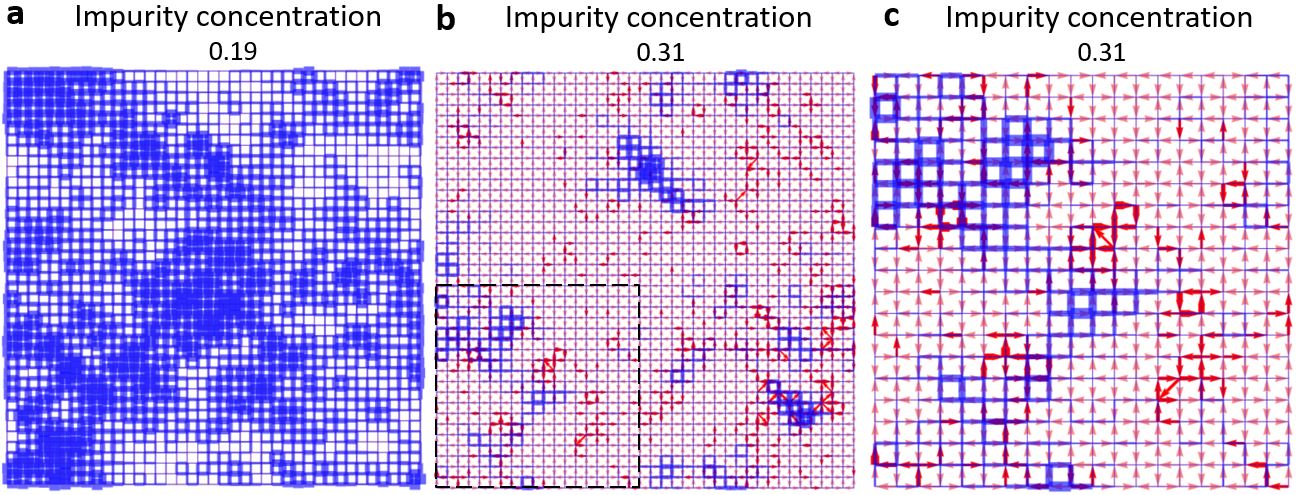}
		\caption{\textbf{Equilibrium current loops}. The blue colored bonds represent the absolute value of the $d$-wave pair field. The red arrows represent the spontaneous generated super-current. The thickness of the arrow denotes the magnitude of the current. The impurity concentration is $0.19$ in panel \textbf{a} and $0.31$ in panel \textbf{b}. No detectable current exists (smaller than $10^{-12}$) in panel \textbf{a} while they are quite apparent for panel \textbf{b}. Panel \textbf{c} is the zoom-in view of the lower-left corner indicated by the black dashed line in panel \textbf{b}. For clarity, only when the magnitude of current is greater than $0.001$ (which is approximately $2/10$ of the maximum current value) do we plot a dark red arrow.  For smaller current values we use  pink arrows to represent it.
		 }
		\label{realspace2}
	\end{center}
\end{figure}

\subsection{Equilibrium current loops}

A subtle but remarkable feature of the mean field solution in the highly overdoped regime is shown in \Fig{realspace2}.  Here, the blue colored bonds represent $|\Delta_{ij}|$ while the red arrows represent
equilibrium supercurrents. Here, the current operator on bond $\avg{ij}$ is given by
\begin{equation}
J_{ij} = it_{ij}\sum_\s\langle c_{i\s}^\dagger c_{j\s} -h.c.\rangle
\end{equation}
We stress that we have not added any time-reversal symmetry breaking perturbation.  The currents form loops, thus satisfying the continuity equations, and are manifestations of the
 spontaneous breaking of time reversal symmetry expected from the random-in-sign Josephson couplings that emerge when superconducting islands are small and sparse.
 The patterns of currents are analogous to  those in an XY spin-glass, with near-degeneracies associated with reversing the local currents around localized loops in different regions of the system.
 This near degeneracy is reminiscent to the existence of two level centers in a glass. They can give rise to orbital paramagnetism. However, since all these effects occur where the superfluid density is small, there may be important qualitative changes in this behavior when the effect of thermal and quantum phase fluctuations are included.
 For instance, the near degeneracy  of the two-level centers can be lifted by tunneling processes in which the local currents reverse direction.
 To the extent that they survive fluctuational effects, these
  spontaneous current loops
  are a qualitatively significant result of the combination of $d$-wave pairing and scalar disorder.

\subsection{Role of the anti-nodal  dispersion}

\begin{figure}[t]
	\begin{center}
		\includegraphics[scale=
		0.44]{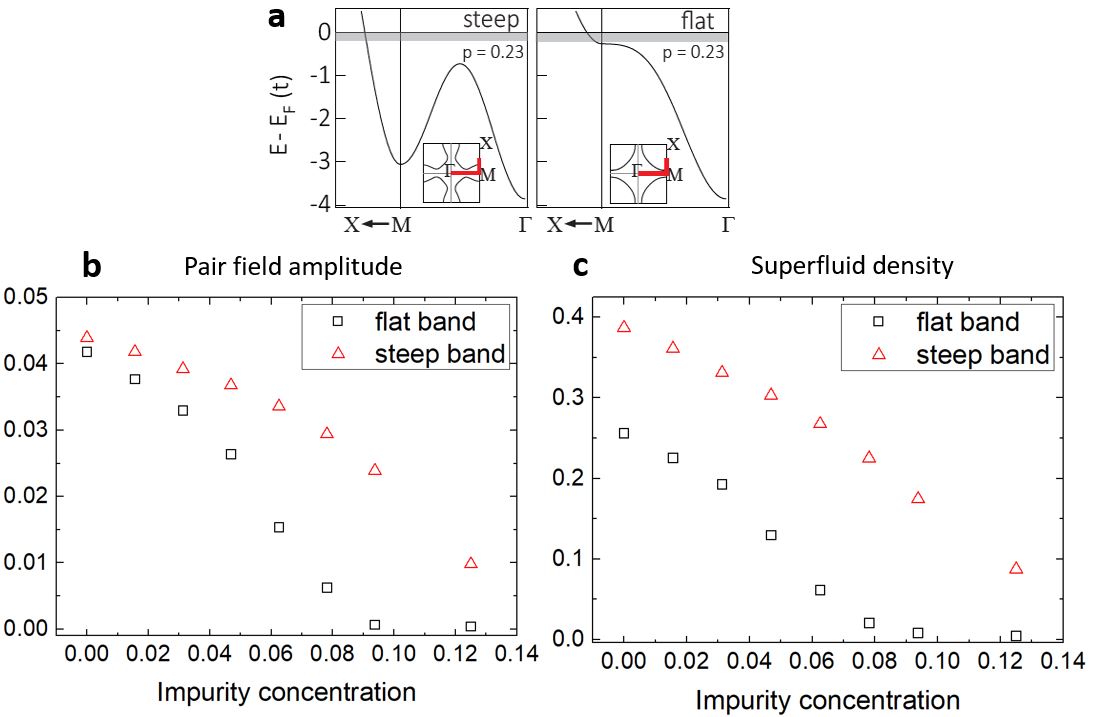}
		\caption{\textbf{The comparison between flat and steep bands}. The spatially averaged zero-temperature d-wave pair field amplitude \textbf{b} and superfluid phase stiffness \textbf{c} computed with the two different band-structures shown in \textbf{a}.  Results for the partially flat band
		are shown as the black squares  and for the steep band by the red.
		}
		\label{flatdisp}
	\end{center}
\end{figure}
Because the d-wave gap is maximal in the anti-nodal region of the Brillouin zone (BZ), many features of the mean-field solution depend sensitively on the band-structure in this region.
Here we show the effects of the flat antinodal dispersion in overdoped cuprates. Such effects have been emphasized in the recent photoemission work\cite{ZXnew}. Due to the enhanced density of states, we find that the existence of
the  flat dispersion amplifies the disorder induced anti-nodal scattering, and hence enhances
 pairing heterogeneity.  Moreover, it leads to a more rapid suppression of
 the zero temperature superfluid density
 with increasing $p$. \\

In \Fig{flatdisp} we compare
results for two band structures: one with a flat dispersion near the antinodes, (as is generic in the cuprates) and the other in which the antinodal dispersion is relatively steep. The
 band structure parameters are chosen such that the doping level is fixed at $23\%$ and
 the Fermi energy ($3.875 t$)
 is the same for the two different band structures (see Supplementary Note 3).  The most significant difference between the flat and steep bands is the existence/absence of a flat dispersion along the BZ boundary
as shown in the relevant part of the BZ in \Fig{flatdisp}a. The spatial averaged zero-temperature pair field amplitudes and superfluid densities as a function of impurity concentration are shown in panel b and c.   Note that in these figures, when the impurity concentration is varied, we  tune the chemical potential so that the hole concentration is unchanged.
It is apparent that the suppression of the pair field amplitude and of the superfluid density are considerably more rapid in the case of flat band. This result is consistent with
  recent ARPES results \cite{ZXnew} and the interpretation therein. We attribute the more rapid suppression of the pair-field amplitude and the superfluid density to the larger density of states associated with the flat band, and the concomitant enhancement of the pair breaking anti-node to anti-node scattering.

\subsection{The specific heat and optical conductivity}

\begin{figure}[t]
	\begin{center}
		\includegraphics[scale=
		0.55]{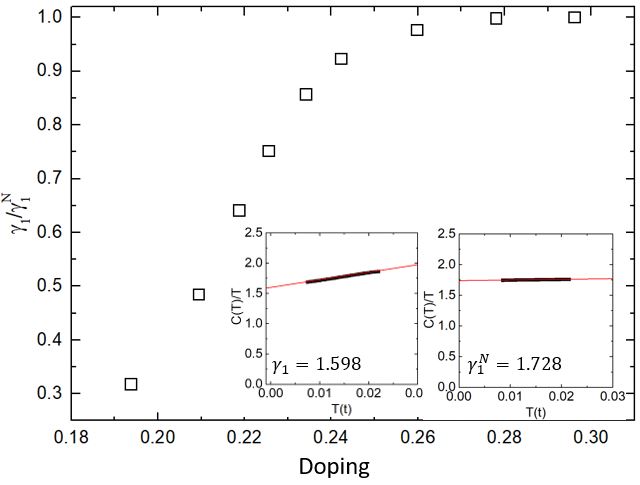}
		\caption{ \textbf{Doping dependence of specific heat coefficient $\gamma_1$ normalized by the corresponding normal-state value $\gamma_{1N}$}. The zero-temperature specific heat coefficient is extracted by fitting $C(T)/T$ using $C(T)/T = \gamma_1 + \gamma_2 T$ and extrapolating to
		$T=0$. The fitting is illustrated in the inset, where the left panel
		is for the d-wave SC state while the right
		corresponds to the normal state.  }
		\label{gamma}
	\end{center}
\end{figure}

A feature of the inhomogeneous state is that there remains a large density of gapless quasi-particle states arising from the approximately normal metallic regions between the superconducting grains.  This is reflected in a residual $T$ linear contribution to the specific heat and to the $\omega \to 0$ optical conductivity that survives even as $T\to 0$.
In \Fig{gamma}  we plot the ratio between the low temperature specific heat coefficient $\gamma\equiv c/T$ and the corresponding value in the normal state as a function of doping concentration. Notably when the doping concentration is high, the ratio approaches one.
 In Supplementary Figure 3 we plot the real part of the optical conductivity as a function of frequency at $T=0$, which shows that a large portion of normal state Drude weight is uncondensed. In Supplementary Note 4, we discuss some discrepancies in the frequency dependence of our result when compared with the experimental data of Ref.\cite{Armitage-2019}. \\

\section{Discussion}

There are several aspects of the superconductor-to-metal transition in overdoped cuprates that have been the subject of various recent theoretical studies.

One issue concerns the microscopic mechanism for the superconductor to insulator transition. While on the underdoped side of the superconducting dome,  the gap scale appears to remain large even as $T_\text{c} \to 0$, on the overdoped side the gap (measured in various ways) decreases significantly as $T_\text{c}\to 0$.  Since it is generally believed that spin-fluctuations are a dominant contributor to the d-wave pairing, and since signatures of incipient antiferromagnetic order become increasingly weak with increasing $p$, it is reasonable to associate the drop in $\Delta$ with a weakening of the pairing interaction.  That such a trend occurs in a Hubbard model with a band structure suitable for the cuprates has recently been shown in Ref. \cite{scalapinoaandmeyer}.  On the other hand, whether the short-range antiferromagnetic correlation drops
sufficiently strongly to account for the demise of the superconducting phase %
is under debate\cite{Bozovic-2013}. In the present study, we showed (at mean-field level) that
even holding the strength of the pairing interaction fixed, a strong drop in the pairing scale can be accounted for simply as a consequence of an increased density of random scattering centers. We consider it likely that both effects play a role in the overdoped cuprates.

Another issue concerns how disorder is treated. In recent theoretical studies 
of overdoped cuprates\cite{scalapinoaandmeyer,brounandhirschfeld,broun} the   effects of disorder are  treated in an effective medium approximation, in which macroscopic fluctuations in the local impurity configurations are averaged out, and the superconducting state is  homogeneous. These studies ignore  any self-organized  granularity.
Nonetheless with suitable choices of parameters, they have been shown to produce phenomenologically reasonable results.  However, we feel that the experimentally  observed\cite{Bozovic-2016}  quantitative similarity between $T_\text{c}$ and $T_\theta$ (i.e. the  $T=0$ superfluid density expressed\cite{Kivelson-1995}  in  temperature units), implicates the reduced superfluid density as the cause of the superconductor to metal transition.  The emergent granularity that we have found provides a theoretically sound origin for such a reduced superfluid density.

The current work, emphasizing the interplay of d-wave pairing and disorder as the cause of a superconductor-to-metal transition,  is qualitatively different from older studies of the superconductor-to-insulator transition
   in the case of an s-wave SC\cite{Scalettar2006,Trivedi2001},
   despite the appearance of self-organized granularity in both cases. In particular,
   the residual resistivity at the end of SC dome is roughly 10 $\mu\Omega$-cm, i.e. two orders of magnitude
   smaller than the quantum of resistivity\cite{Bozovic-2016} per Cu-O plane. Although our model and calculation are similar to some previous works of mean-field calculation on dirty d-wave
   superconductors \cite{dirtydwave1,dirtydwave2,dirtydwave3,dirtydwave4,dirtydwave5,dirtydwave6},
   the issues we address, namely the disorder driven superconductor-to-metal transition,  have not been sharply articulated previously.

Thus, we propose the key for understanding the essence of overdoped cuprates is the combined effect of disorder and $d$-wave pairing. These features,
especially when combined with relatively flat bands in the antinodal regions of the BZ, lead to  a self-organized granular superconducting state.
 As an additional consequence of the $d$-wave pairing, the Josephson coupling between the superconducting islands is generically frustrated in the strong disorder limit. As a consequence there are spontaneous current loops and associated local breaking of time reversal symmetry.
  There are several possible observable signatures of this\cite{Kivelson-2015};  for instance, we find that under conditions in which the mean-field zero temperature superfluid density  is significantly suppressed by disorder,
  the superfluid density can be an {\it increasing} function of an applied field for small fields. In addition, due to the formation of local current loops, the system
  can exhibit a paramagnetic Meissner effect\cite{Wohlleben-1992}, namely,
  a paramagnetic response upon applying
  a small magnetic field. Such effects are beyond the reach of theories which treat the disorder in effective medium approximation.

Finally,
we reiterate that  the model, and the mean-field treatment of it we have discussed,
are simplified  compared to any actual cuprate.  Indeed, it follows  from the present results that thermal and quantum fluctuations,
not included explicitly in our treatment so far, are inevitably important in the vicinity of superconductor-to-metal transition.
 Thus, while our work is a step towards understanding the basic physics of superconductor-to-metal transition in overdoped cuprates,
 many important issues are still unsettled.

\section*{Methods}
\subsection* {\bf Self-consistency mean-field calculation}
\noindent {The results are obtained by numerically self-consistency mean-field calculation. The system size in our calculation is $40\times40$. We average over distinct disorder configurations to compute the physical observables. Typically, the number of disorder configurations is $64$, but we average over $128$ configurations when the impurity concentration is large and the superconducting pairing is highly inhomogeneous.}

\section*{DATA AVAILABILITY}
\noindent The data that support the findings of this study are available from the corresponding author upon reasonable request.

\section*{Acknowledgements}
We would like to thank D.J. Scalapino, Yu He, and Zhi-Xun Shen for helpful discussions, and S. Uchida and T. Maier for constructive criticism. This work was primarily funded by the US Department of Energy, Office of Science, Office of Basic Energy Sciences, Materials Sciences, and Engineering Division under Contract No. DEAC02-05-CH11231 (Quantum Material Program KC2202). The computational part of this research is supported by the US Department of Energy, Office of Science, Office of Advanced Scientific Computing Research, Scientific Discovery through Advanced Computing (SciDAC) program. This research is funded in part by the Gordon and
Betty Moore Foundation. S.A.K. was supported in part by the U. S. Department of Energy (DOE) Office of Basic Energy Science, Division of Materials Science and Engineering at Stanford under contract No. DE-AC02-76SF00515 at Stanford.

\section*{Author Contributions}
D.H.L. and S.A.K. conceived the project. Z.X.L. performed the calculations. All the authors contribute to editing the manuscript.

\section*{Competing Interests}
The authors declare no competing financial or non-financial interests.

\begin{widetext}
\section{Supplementary Materials}

\renewcommand{\theequation}{S\arabic{equation}}
\setcounter{equation}{0}
\renewcommand{\thefigure}{S\arabic{figure}}
\setcounter{figure}{0}
\renewcommand{\thetable}{S\arabic{table}}
\setcounter{table}{0}

\renewcommand{\figurename}{{\bf Supplementary Figure}}
\renewcommand{\tablename}{{\bf Supplementary Table}}
\renewcommand{\thetable}{\arabic{table}} 

\subsection{Supplementary Note 1: The zero-temperature pair field amplitude and superfluid density\\}
We employ self-consistent, superconducting, mean-field approximation to study Eq.(1) in the main text. We consider two different forms of $H_{\rm int}$: 1) To trigger d-wave pairing, we use the nearest-neighbor Heisenberg antiferromagnetic exchange interaction $ H_{\rm int}=J\sum_{\<ij\>}\v S_i\cdot\v S_j $ . 2) To trigger s-wave pairing, we use the attractive Hubbard interaction $H_{\rm int}=-U\sum_{i}c^\dagger_{i\uparrow}c_{i\uparrow}c^\dagger_{i\downarrow}c_{i\downarrow}$. After the mean-field decoupling, we obtain, respectively,
\be
H^{\rm d}_{\rm MF} &=& -\frac{1}{2}\sum_{ij} \Delta_{ij} (c^\dagger_{i\uparrow}c^\dagger_{j\downarrow}-c^\dagger_{i\downarrow}c^\dagger_{j\uparrow}) + h.c. \nonumber\\
H^{\rm s}_{\rm MF} &=& - \sum_{i} \Delta_{i} c^\dagger_{i\uparrow}c^\dagger_{i\downarrow} + h.c.
\ee
where pair field for d-wave and s-wave are given by: $\Delta_{ij} = J \avg{c_{j\downarrow}c_{i\uparrow}-c_{j\uparrow}c_{i\downarrow}}$ and $\Delta_i = U \avg{c_{i\downarrow}c_{i\uparrow}}$, respectively, which are determined self-consistently.\\

To study the averaged pair field amplitude and superfluid density in strongly disordered regime,  we plot the logarithm of the pair field amplitude and superfluid density as a function of doping concentration, or impurity concentration (see Fig.2 in main text) in \Fig{logplot}.

\begin{figure}[h]
	\begin{center}
		\includegraphics[scale=0.55]{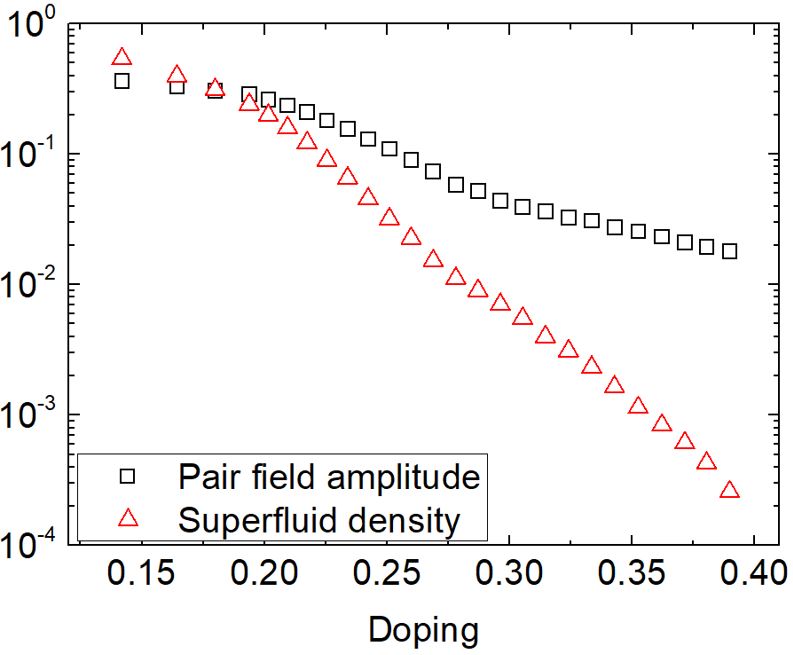}
		\caption{\textbf{The logarithmic plot of pair field amplitude and superfluid density as a function of doping concentration for the d-wave pairing}.  }
		\label{logplot}
	\end{center}
\end{figure}

In the regime where doping (hence impurity) concentration is high, the pair field amplitude obeys approximately an exponential behavior. This is consistent with the prediction of Ref.\cite{Kivelson-2018}, suggesting the significant pair field originates from the exponentially rare regions  in which the disorder is relatively weak. Compared with pair field amplitude, the superfluid drops considerably more rapidly with increasing doping concentration, or impurity concentration.

Assuming the presence of 90 degree rotation symmetry after impurity averaging, we compute the zero temperature superfluid density by twisting the phase of the pair field in a pre-chosen, say, x-direction. Alternatively, it can be calculated as the current response to a gauge field in $\hat{x}$. Using the standard Kubo formula:
\be
\rho_s = -\avg{K_x} - \Lambda_{xx}(q_x=0,q_y \rightarrow 0, \omega = 0).
\label{S2}
\ee
Here $$K_x=\frac{1}{N}\sum_{i,\sigma}\left(-t_1 c^\dagger_{i\sigma}c_{i+x \sigma} - t_2 c^\dagger_{i\sigma}c_{i+x+y \sigma} - t_2 c^\dagger_{i\sigma}c_{i+x-y \sigma} +  h.c.\right),$$ with $t_1$ being the hopping integral between nearest-neighbor sites, and $t_2$ is the hopping integral between second-neighbor sites, $\Lambda_{xx}(\v q,\omega)$ is the current-current correlator defined as follow:
\be
\Lambda_{xx}(\v q,\omega) = \frac{i}{N}\int^{\infty}_0 dt e^{i\omega t} \avg{[J_x(t,\v q),J_x(0,-\v q)]}.
\label{S3}
\ee
In the equation above $N$ is the number of lattice sites, $J_x(\v q) = \sum_i J_x(\v r_i) e^{i \v q\cdot\v r_i}$ is the Fourier transform of the current operator along x-direction.\\

\begin{figure}[h]
	\begin{center}
		\includegraphics[scale=0.75]{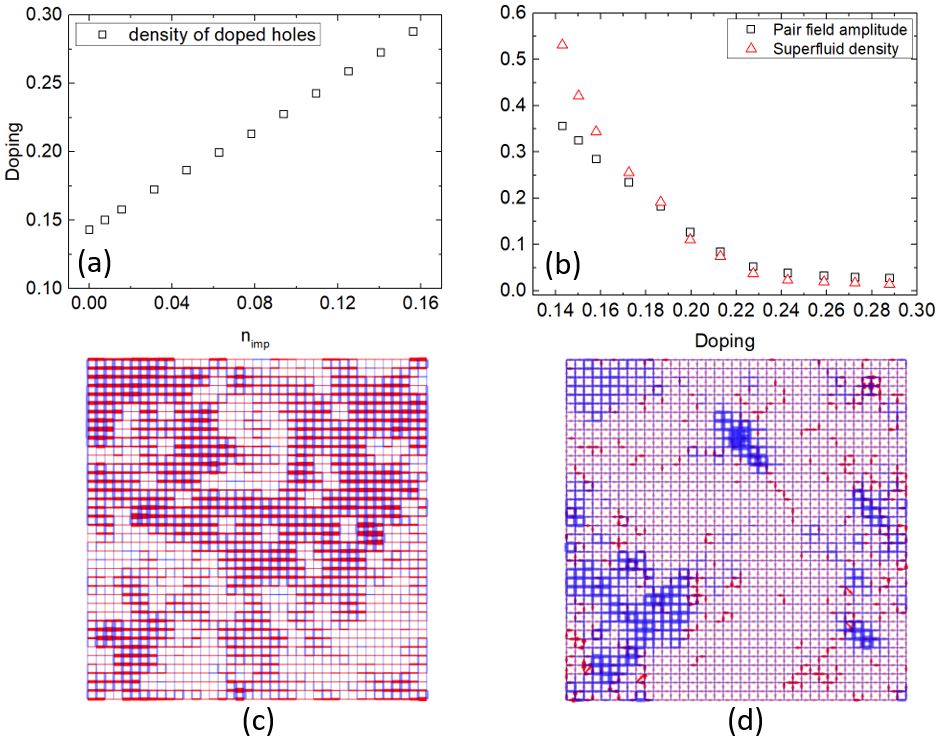}
		\caption{\textbf{The results of the zero-temperature averaged pair field amplitude, superfluid density, and spontaneous current loops, for disorder strength $w=3$}. (a) The pair field amplitude and superfluid density as a function of doping level. (b) Real-space distribution of the pair field for impurity concentration $n_{\rm imp}=0.078$: the thickness of the symbols represents the magnitude and the color(red positive, blue negative) represents the sign. (c) Real-space distribution of the spontaneous current loops and the pair field amplitude for impurity concentration $n_{\rm imp} = 0.125$: the blue colored bonds represent the pair field amplitude, and the red arrows represent the spontaneous generated super-current.  }
		\label{resultw3}
	\end{center}
\end{figure}

\subsection{Supplementary Note 2: The pair field amplitude and superfluid density for disorder strength $w=3$}

In the main text, we report the averaged zero-temperature pair field amplitude and superfluid density for a fixed disorder strength $w=1$. In this section, we present the results for a different value of disorder strength $w=3$,  to demonstrate that the main qualitative results do not depend on the disorder strength. We choose the uniform chemical potential $\mu = -1.08$ such that the density of doped holes is $p= 0.145$, when the system is disorder free. After fixing $\mu$, doping level increases monotonically with impurity concentration $n_{\rm imp}$. The dependence of the density of doped holes, $p$, on the impurity concentration $n_{\rm imp}$ is shown in \Fig{resultw3}(a). We calculate the averaged zero-temperature pair field amplitude and superfluid density as a function of doping level, as shown in \Fig{resultw3}(b). Like the case of $w=1$, the impurity concentration is large, the superconducting pair field amplitude becomes highly heterogeneous (shown in \Fig{resultw3}(c)). In addition, like the case of $w=1$, spontaneous current loops emerge when the disorder is strong (see \Fig{resultw3}(d)). These findings are consistent with the results when disorder strength is $w=1$.

\subsection{Supplementary Note 3: The band structures of  the``flat band'' and ``steep band'' models}

In the main text, we present the mean-field results of the averaged zero temperature pair field amplitude and superfluid density for the ``flat'' and ``steep''  band structures (see Fig. 6). The parameters used to produce these band structures shown are given as follows: i) For the ``flat band'': $t_1 = 1$, $t_2 = -0.05$, $t_3 = 0.2$ and $\mu = -0.778$. ii) For the ``steep band'': $t_1 = 1$, $t_2 = -0.4$, $t_3 = 0.68$ and $\mu = -1.263$. Here $t_1$, $t_2$ and $t_3$ are the hopping integrals between nearest-neighbor sites, second-neighbor sites and third-neighbor sites, respectively, and $\mu$ is uniform chemical potential.  Under the choice of these parameters, the doping level (23\%) and the overall band width  (3.875) are the same for the two different band structures.  Note the results reported in Fig. 6 (b) and (c) are obtained under a fixed doping level. This is achieved with increasing impurity concentration, by fine tuning the value of uniform chemical potential.

\begin{figure}[h]
	\begin{center}
		\includegraphics[scale=
		0.8]{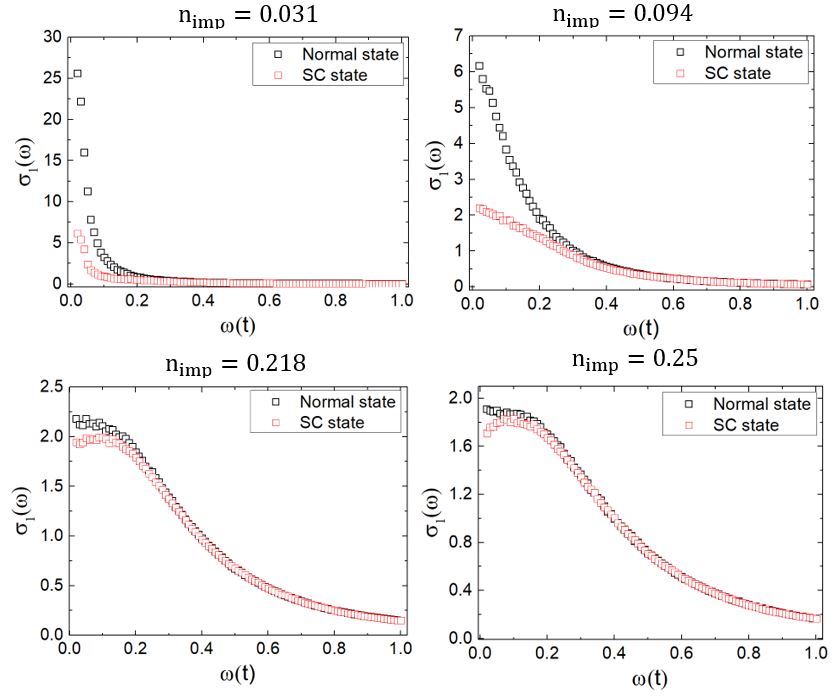}
		\caption{ \textbf{Frequency dependence of the real part of optical conductivity at zero temperature for different impurity concentrations}. The normal state results are obtained by setting the pair field to zero.  }
		\label{optical}
	\end{center}
\end{figure}

\subsection{Supplementary Note 4: The optical conductivity}

In this section, we present the zero-temperature optical conductivity for different impurity concentrations. The optical conductivity is computed from the current-current correlator defined as \Eq{S3}:
\be
\sigma(\omega) = \frac{1}{i\omega}\Lambda_{xx}(\v q=0,\omega)
\ee
The frequency dependence of the real-part of $\sigma(\omega)$  is shown in \Fig{optical} for several  impurity concentrations. The normal state results are obtained by setting the pair field to zero. The result clearly shows that when the impurity concentration is high, a large fraction of the normal-state Drude weight is uncondensed. The result is consistent with physical picture that we have superconducting islands embedded in a normal metal matrix. However, when impurity concentration is high, our optical conductivity results show a broad peak at finite frequency. The peak position is consistent with the averaged superconducting gap. This feature is not observed in the experimental measurement\cite{Armitage-2019}. A possible origin of this discrepancy is that the experimental measurements have not yet reached the frequency regime where the averaged gap will be manifested. Another possibility is the fact that these peaks are very broad, making it difficult to observe experimentally. In any case a more thorough analysis of this discrepancy is left in future studies.

\begin{figure}[h]
	\begin{center}
		\includegraphics[scale=
		0.8]{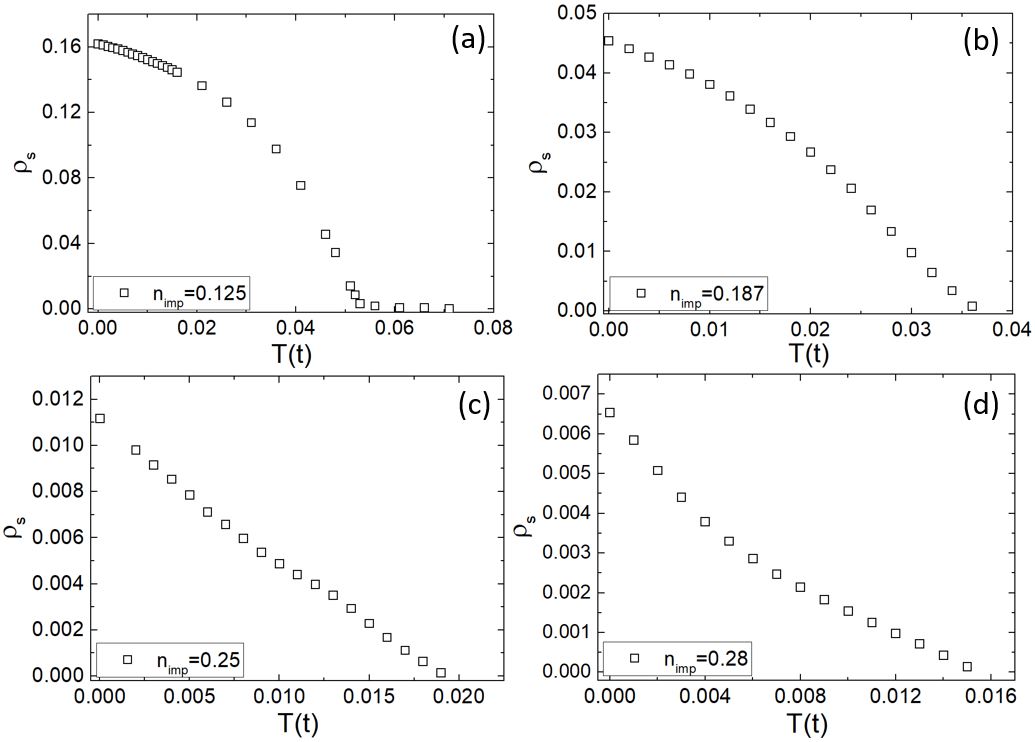}
		\caption{ \textbf{The temperature dependence of superfluid density for different impurity concentrations}. We use the unit in which the Boltzmann constant $k_B = 1$. This results only include the quasiparticle effects. The thermal phase fluctuations are omitted. }
		\label{superfluiddensity}
	\end{center}
\end{figure}

\subsection{Supplementary Note 5: The superfluid density at non-zero temperature}

In this section, we present the results of superfluid density at non-zero temperature for four different impurity concentrations. Because the present calculation neglects the suppression of superfluid density by thermal phase fluctuations,  it can only give reasonable results at low temperature. The temperature dependences of superfluid density for four different  impurity concentrations are shown in \Fig{superfluiddensity}. For sufficiently low temperature, we find the superfluid density drops approximately linearly with temperature. This behavior is consistent with the experimental results for LSCO\cite{Bozovic-2016}.  However, in the same experiment, the linear decrease of the superfluid density with temperature is obeyed all the way up to $T_{\text c}$. Clearly, this is not something we can explain at present.

\end{widetext}

\end{document}